\documentclass[table]{article}

\usepackage{graphicx}
\usepackage{paralist}
\usepackage[numbers,sort&compress]{natbib} %
\usepackage{url}
\usepackage{ctable}
\newcommand{\mytoprule}{\specialrule{\heavyrulewidth}{\aboverulesep}{0em}}
\newcommand{\mymidruletop}{\specialrule{\lightrulewidth}{\aboverulesep}{0em}}
\newcommand{\mymidrulebottom}{\specialrule{\lightrulewidth}{0em}{\belowrulesep}}
\newcommand{\myvertsp}{{\Large$\phantom{\mid}$\hspace{-.25em}}}
\newcommand{\myhlcell}{\cellcolor{lightgray}\myvertsp}

\makeatletter
\begin{document}
\title{A Manifesto for Applicable Formal Methods\thanks{This preprint has not undergone peer review (when 
applicable) or any post-submission improvements or corrections. The Version of Record of this 
article is published in Software and Systems Modelling, and is available online at \texttt{https://doi.org/10.1007/s10270-023-01124-2}.}}

\author{Mario Gleirscher$^1$ $\cdot$ Jaco van de Pol$^2$ $\cdot$ Jim Woodcock$^3$ %
  \\
 \normalsize
 $^1$\texttt{gleirscher@uni-bremen.de}, University of Bremen, Germany\\
 \normalsize
 $^2$\texttt{jaco@cs.au.dk}, Aarhus University, Denmark\\
 \normalsize
 $^3$\texttt{jim.woodcock@york.ac.uk}, University of York, United Kingdom
}

\maketitle  
\begin{abstract}
  Formal methods were frequently shown to be effective and, perhaps because
  of that, practitioners are interested in using them more often.
  Still, these methods are far less applied than expected, particularly, in critical domains where they are strongly recommended and where they have the greatest potential.
  Our hypothesis is that formal methods still seem not to be applicable enough or ready for their intended use.
  In critical software engineering, what do we mean when we speak of a \emph{formal method}?  And what does it mean for such a method to be \emph{applicable} both from a scientific and practical viewpoint?
  Based on what the literature tells about the first question, with
  this manifesto, we lay out a set of principles that when followed by
  a formal method give rise to its mature applicability in a given
  scope.
  Rather than exercising criticism of past developments, this
  manifesto strives to foster an increased use of formal methods to
  the maximum benefit.
\end{abstract}

\section{Introduction}
\label{sec:introduction}

Formal methods~(FMs) have been an active research area for decades.
Theoretical foundations~\citep{HuthR2004}, method
applications~\citep{Aichernig2003,Gnesi2013,Boulanger2012}, as well as
effective ways to transfer~\citep{Miller2010,OHearn2018} them to the
practising engineer have been thoroughly discussed and empirically
evidenced~\citep{Sobel2002,Pfleeger1997-Investigatinginfluenceformal}.
The resources to learn about 
these methods  range from early syllabuses~\citep{Garlan1992-Formalmethodssoftware} to recent course
materials,\footnote{Available from the Formal Methods Europe association:
  \url{https://fme-teaching.github.io/courses}}
tutorial papers~(e.g.,~\cite{Behrmann2004-TutorialUppaal}), tool
manuals,\footnote{E.g.,~\url{http://www.prismmodelchecker.org/manual}}
text books~(e.g.,~\citep{Jones1980-Softwaredevelopment}, \cite{Nipkow2014-ConcreteSemantics}), and a
community wiki.\footnote{The Formal Methods Body of Knowledge (FMBoK):
  \url{https://formalmethods.wikia.org}}  However, evidence on successful formal method teaching, training,
and teaching-based transfer is still missing
\citep{Garavel2020-2020ExpertSurvey,Gleirscher2020-FormalMethodsDependable}.

Driven by the inspiration and critique of expert voices from
academia~\citep{Hall1990,Bowen1995,Knight1997,Barroca1992,McDermid1998-Towardsindustriallyapplicable,Parnas2010} and
industry~\citep{Abrial2006-Formalmethodsindustry}, formal methods are
considered to be one of the most promising tools to develop highly
dependable software for critical applications
\citep{Garavel2020-2020ExpertSurvey}.  Developers of formal methods
have always aimed at applicability in practical contexts, notably with
different degrees of success.
Indeed, many practitioners believe in the high potential of such
methods and would use them to their maximum benefit, whether directly
or through powerful software
tools~\citep{Gleirscher2020-FormalMethodsDependable}.  Although, there
is wide interest in applying these methods in the engineering practice
of dependable systems and software, this domain has not yet
successfully adopted formal methods.  It is observed
(e.g.,~\cite{Garavel2020-2020ExpertSurvey,Gleirscher2020-FormalMethodsDependable})
that their use is still significantly weaker than expected, most
alarmingly, even in critical
domains~\citep{Gleirscher2019-NewOpportunitiesIntegrated} where their
application is, in parts and through a wide range of standards~(e.g.,~IEC~61508 and 62443, DO-178), strongly recommended.

It is thus reasonable to assume that FMs (still or again) seem not to be applicable enough (or ready) for their intended purpose.  An alternative explanation would be that modern programming languages and environments implicitly support a good part of what would have been called formal development in the period from the 1970s to the 1990s and avoid many of the hard sought-after errors, FMs were originally supposed to unveil.  This explanation is, however, only reasonable if we ignore the massive increase in software and hardware complexity since then and the increase in use of software in critical areas.  Consequently, new kinds of problems and errors have shown up and the original justification for the use of FMs remains valid, albeit at different levels of abstraction. 

In that light, the beneficial use of formal methods is hindered, for example, by poor scalability, missing or inadequate tools, scarce teaching and training, and thus a lack of trained personnel \citep{Gleirscher2020-FormalMethodsDependable}.  The lack of recent knowledge about these obstacles and the effectiveness and productivity of formal methods \citep{Gleirscher2019-NewOpportunitiesIntegrated} raise a high demand for formal method research and goal-directed collaborations between academia, regulators, and industry.
To help research and transfer efforts gain momentum and foster
success, we
suggest some guiding
principles of applicable formal methods in form a manifesto.

\paragraph{Outline.}

The Sections~\ref{sec:background} and~\ref{sec:relwork} provide the background and motivation of this manifesto and highlight related work.
Section~\ref{sec:manifesto} presents the manifesto and its principles.
Section~\ref{sec:success-stories} highlights several formal methods success stories.
Section~\ref{sec:implications} summarises aims, suggests actions to
implement the manifesto, and outlines expected impacts of these actions.
Section~\ref{sec:inaction} warns about potential consequences of inaction by the community,
and
Section~\ref{sec:conclusions} concludes.

\section{Background}
\label{sec:background}

\paragraph{What is a ``formal method''?}
\label{sec:what-formal-method}

There are many useful characterisations available from the literature.  For example, 
the \emph{IEEE~Software Engineering Body of Knowledge} says: ``formal methods are software engineering methods used to specify, develop, and verify the software through application of a rigorous mathematically based notation and language'' \citep[p.~9-7, Sec.~4.2]{Bourque2014-GuideSoftwareEngineering}.
This recent definition~\citep{Baudin2021-doggedpursuitbug} covers the relevant
aspects quite well, stating that ``formal methods are a set of
techniques based on logic, mathematics, and theoretical computer
science which are used for specifying, developing, and verifying
software and hardware systems.''
We slightly refine this notion, saying that, by a \emph{formal
  method}, we refer to an explicit mathematical model and sound
logical reasoning about critical
properties~\citep{Rushby1994-Criticalsystemproperties}---such as
reliability, safety, security, dependability, performance,
uncertainty, or cost---of a class of electrical, electronic, and
programmable electronic or software systems.  Model checking, theorem
proving, abstract interpretation, assertion checking, and formal
contracts are classical examples of versatile formal methods.

\paragraph{What makes formal methods so special?}

Generally, a method can be thought of as a step-wise recipe providing
guidance for its user regarding the next steps to take in certain
situations.  In analogy to other engineering disciplines, a formal
method pushes the role of mathematics and logic in software
engineering to
\begin{itemize}
\item make objects explicit (e.g.,~natural processes, 
  information, peoples' thoughts) through
  \emph{notation} with a precise \emph{meaning} agreed within a
  domain,
\item reduce ambiguity about or subjective interpretation of
   these objects (e.g., a system or its functioning) and foster a precise understanding of that domain, and
\item support mechanisation of critical or tedious tasks (e.g.,~analysis,
  verification).
\end{itemize}
These features enable one to distinguish formal from informal (or
``non-formal'') methods.\footnote{For example, UML is a standardised
  notation and carries, through its many historically-inspired
  concepts, flavours of methods.  However, to obtain a corresponding
  FM, UML's concepts need to be underpinned with a precise semantics and
  a method to construct and reason about UML models.
  Similarly, Java has a fairly well specified grammar and platform.
  However, to obtain an FM for Java, executions of a Java program need
  to be given a precise semantics as well as a method to reason about it
  logically and conceptually.}  A formal method is more than 
a (modelling) notation or a development or analysis method and
different from a programming language or a software engineering tool.

Although there is no clear boundary between formal and informal
methods (formality may occur in degrees), one can be of the opinion
that a software engineering method is ``informal'' if the use of
mathematics and logic is neither essential nor required to
create reliable results, and it is ``formal'' otherwise.

\paragraph{What do we mean by ``applicability''?}

Generally, ``applicable'' means \emph{capable of being applied}, within some defined and practically relevant scope.  More specifically, \emph{applying a formal method} involves its \emph{use} in the design, development, and analysis of a critical system and its substantial \emph{integration} with the used development methodologies~(e.g., structured development, model-based engineering, assertion-based programming, test-driven development), specification and modelling notations (e.g., UML, SysML), programming languages, and tools.  When we use the terms ``applicable'' or ``applicability'', we refer to a desirable degree or \emph{level of maturity}\footnote{In analogy to NASA's Technology Readiness Levels (NASA) and the CMMI framework from CMU.} of a formal method.  Consequently, this notion suggests some quantitative (e.g., performance or economic) assessment to be able to make objective statements about the level of maturity and, thus, the applicability of a formal method.

\paragraph{When do we expect a formal method to be applicable?}

We need applicability whenever we suggest a FM as a critical (quality assurance) \emph{instrument} to be used in a critical engineering \emph{task}.  That task will primarily be a practical software engineering task but it can also be an engineering task in computer science research and teaching.

We need applicability if the expected \emph{benefit} from using a
formal method in a task~(e.g., early error reduction, design improvement, didactic gain, scientific insight) justifies the expected \emph{cost of
  applying} it (e.g., formalisation effort, time, and resources) but it
does not justify the \emph{cost of not applying} it (e.g., late failure
handling costs, failure consequences).

\paragraph{What makes formal method applicability so special?} 

What makes it different from applicability of other modelling or programming methods, techniques, or languages?  A formal method requires one to use (through tools and with guidance) mathematical structures to represent and make concise the meaning of objects (e.g., software or system behaviour, data sets) to be reasoned upon.  The proper understanding and efficient use of such structures needs special abstraction capabilities, mathematical skills to be taught, and continuous application-oriented training.  An applicable formal method is a method that addresses these very specific requirements in this particular context.

\paragraph{What is a manifesto and why do we need one?}
\label{sec:motiv}

A manifesto can be understood as ``a series of technical or expert
views on a particular engineering task''
\citep{Ladkin2018-CriticalSystemAssurance}, ``a set of commitments''
of a community~\citep{Rae2020-manifestoRealitybased}, or ``a focal
point of reference'' catalysing
communities~\citep{Becker2015-SustainabilityDesignSoftware}.  Inspired
by successful similar efforts in other
domains~\citep{Becker2015-SustainabilityDesignSoftware},\footnote{See
  also the GNU Manifesto~(1985,
  \url{https://www.gnu.org/gnu/manifesto.html}) and the Agile
  Manifesto~(2001, \url{http://agilemanifesto.org}).} we summarise: A
manifesto expresses consensual agreement among
stakeholders~(e.g., experts, thought leaders, users) in a domain, it
is based on corresponding definitions, it concisely %
conveys
guidance in terms of principles, it discloses aims and commitments in
form of an appeal, it suggests actions, and it can join forces and, thus, initiate change.

\section{Related Work}
\label{sec:relwork}

Our manifesto can be seen as a specific supplement of the \emph{Verified Software Initiative}~\citep{Hoare2009-verifiedsoftwareinitiative}, which has the long-term aim to perform wide-ranging verification experiments and case studies, improve the tool landscape, and foster transfer of FM research to industry. Ladkin's manifesto~\citep[Ch.~10]{Ladkin2018-CriticalSystemAssurance} includes principles and steps of how formal methods could be used in practical and standard-compliant software assurance.  While his manifesto covers many areas of software assurance, the section on FM guidance concentrates on the use of FMs in assurance.  Our manifesto complements Ladkin's with guidance on how to prepare FMs to be applicable in assurance and beyond.  Rae et al.'s manifesto~\citep{Rae2020-manifestoRealitybased} aims at an improvement in the use of research methods in safety science, not touching on FM applicability in software safety.

\section{The Manifesto and Its Ten Principles}
\label{sec:manifesto}

We present the manifesto with its goals, principles, and aims concisely in Table~\ref{tab:principles} and then explain and comment on each principle in more detail.

\begin{table}[t]
    \caption{The manifesto and its ten principles at a glance}
    \label{tab:principles}
    \begin{tabularx}{\textwidth}{>{\bfseries}lX}
    \mytoprule
    \multicolumn{2}{l}{\myhlcell\textbf{Motivation}}
    \\\mymidrulebottom
      Success stories
      & FMs aim at applicability and were shown to be effective. \\
      Visible demand
      & Practitioners are interested in using FMs more frequently. \\
      But: Scarce use
      & FMs are far less applied than expected, particularly, in critical domains. \\
    \mymidruletop
    \multicolumn{2}{l}{\myhlcell\textbf{Diagnostic Finding}}    
    \\\mymidrulebottom
    \multicolumn{2}{p{.97\textwidth}}{Formal methods (still) seem not to be
      applicable enough (or ready)
      for their intended use.}
    \\\mymidruletop
    \multicolumn{2}{l}{\myhlcell\textbf{Ten Principles / Precepts / Commitments}}
    \\\mymidrulebottom
      Scope
      & Clearly define the scope of applicability \\
      Methodology
      & Provide concepts, tools, and procedural guidance (for scalability)\\
      Integration
      & Integrate with methods, modelling techniques, and prog. languages \\
      Explainability
      & Allow established claims to be communicated precisely and clearly\\
      Automation
      & Provide automated abstractions (for scalability)\\
      Scalability
      & Be applicable to the size/complexity of systems operated in practice\\
      Transfer
      & Provide teaching and training strategies\\
      Usefulness
      & Provide evidence on effectiveness (for a good cost/benefit ratio)\\
      Ease of Use
      & Provide evidence on efficiency (for a good cost/benefit
        ratio)\\
      Evaluation
      & Demonstrate applicability in a credible way
    \\\mymidruletop
    \multicolumn{2}{l}{\myhlcell\textbf{Aims, Actions, and Expected Impacts}}    
    \\\mymidrulebottom
      \multicolumn{2}{p{.97\textwidth}}{
      \textbullet~\textbf{Provide guidance for} performing, writing, and
        reviewing formal method \textbf{research}\newline
        \textbullet~\textbf{Drive} the selection of relevant unsolved (benchmark or fundamental) \textbf{challenges} 
        and \textbf{stimulate research proposals} \newline
       \textbullet~\textbf{Foster interactions} between academia and industry \newline
       \textbullet~\textbf{Establish connections} between formal method developers and users (through
        explainability) and customers (through economical arguments)} 
    \\
    \bottomrule
    \end{tabularx}
\end{table}

\paragraph{The Ten Principles of Applicable Formal Methods.}
\label{sec:princ-appl-form}

In order to evidence \emph{applicability} both in research and in
practical software engineering, a formal method should ideally implement
all of the following principles.

\begin{description}
\item[Scope] It should clearly define its scope of
  applicability\footnote{For example, embedded software engineering
    research or safety-critical software practice in the automotive
    control domain.}, its domain specificity, and come with comprehensible guidance on how it is to be
  applied within that given scope.  The restriction to a limited scope
  can reduce the complexity of the formal model and, thus, increase
  \emph{Ease of Use} and support other principles.
  
\item[Methodology] It should provide a step-wise recipe, procedural
  guidance for method users regarding possible next steps to be taken
  in corresponding situations.  For example, it should support
  composition, modularity (e.g., using formal reasoning  \cite{Collette2000-EnhancingTractabilityRelyGuarantee} 
  about contracts \cite{Meyer1992-Applyingdesigncontract}), 
  and refinement, and come with a variety of
  sound abstraction or simplification techniques.

\item[Integration] It should create benefits through integration with
  other methods.  For example, it should be
  integrated\footnote{Conceptually aligned, representing a semantic
    layer for another method, represented to the user through a common
    tool layer.} with
  \begin{inparaenum}[(i)]
  \item an established formal method or 
  \item a widely-used modelling technique (e.g.,~UML State Charts), 
  \item programming language (e.g.,~Java), or
  \item process model (e.g.,~Scrum).
  \end{inparaenum}
  Integration in this way is supposed to increase \emph{Usefulness}
  and \emph{Ease of Use}.

\item[Explainability] After a successful application of a formal method, it should
be clear what has been demonstrated. A minimal requirement is that it can be stated
precisely which claim has been established (as in a mathematical theorem). A stricter
requirement is that a certificate can be generated, which enables checking the claim
independently. Last but not least, it requires that the claim (including the
underlying modeling assumptions) can be communicated to human domain experts and
maybe even to end users.  Explainability in this way is supposed to
increase \emph{Usefulness}.
  
\item[Automation] It should come with tool support that prevents its
  user from tedious work steps and helps them to focus on essential
  and creative steps.  In particular, it should provide automation
  support for any obvious/useful abstraction required to be crafted to
  apply the method to the maximum benefit.
  Automation usually pertains to difficult or tedious tasks and can,
  thus, increase \emph{Scalability} towards industrial-sized systems.
  
\item[Scalability] It should be applicable at a practically relevant
  scale,\footnote{Where scale may be quantified as, for example,~lines of code,
    the number of fulfilled requirements or discharged theorems, the
    size of a state space, or by a measure of complexity.}  manageable
  with reasonable effort as a function of that scale.
  This principle is likely to be fostered by a clear
  \emph{Methodology}~(e.g.,~superior algorithms, abstraction, modular
  approaches) and strong \emph{Automation}.

\item[Transfer] It should be accompanied with a teaching and training
  strategy and corresponding materials.%
  \footnote{Educational prerequisites, theoretical background material, examples, case
    studies, user guides, tool manuals.} This strategy and the materials may differ from one formal method to another.  However, also average graduate students
    and experienced engineers should be able to learn and apply a method with reasonable effort. 
  
\item[Usefulness] Its effectiveness should be evidenced.  For example,
  it should be demonstrated (e.g., by means of case studies or
  controlled experiments) what would have been different if a
  conventional or non-formal alternative had been used instead (e.g.,
  through comparison of relative fault-avoidance or fault-detection
  effectiveness and the economic impacts of these metrics).
  Usefulness as the governing factor for applicability will be a
  result of other principles, such as \emph{Explainability}.

\item[Ease of Use] It should be efficiently\footnote{Note that the term ``efficiency'' 
here refers to the gain/effort ratio on the user's side.  Efficiency in terms of short tool run-time or low algorithmic complexity is subsumed under \emph{Automation}, \emph{Scalability}.} applicable.  For example,
  it should provide concepts, abstractions, or modelling and reasoning
  primitives that help users with appropriate skills (cf.~\emph{Training}) to apply it with reasonable effort
  (e.g., low abstraction effort, low proof complexity, high
  productivity) within the specified scope.  Ease of use will be a
  result of other principles, such as \emph{Scalability} and \emph{Automation}.
  \emph{Usefulness} and \emph{Ease of Use} refer to the two main constructs of the \emph{Technology Acceptance Model}~\citep{Davis1989-PerceivedUsefulnessPerceived}, a widely used model for the assessment of end-user information technology.

\item[Evaluation] It should demonstrate its applicability in a
  credible way (e.g., with representative examples, with tools usable
  by other researchers or practitioners) that it is applicable to the
  range of engineering problems and systems in its specified scope. It
  should provide information about both its benefits and foreseen
  challenges, limitations, or barriers when applied.  This principle
  integrates the scientific method into the argumentation of FM
  applicability.
\end{description}

\section{Success Stories of FM Integration and Transfer}
\label{sec:success-stories}

There is plenty of anecdotal and stronger evidence on applicable formal methods, not least in the form of success stories of research integration, application, and transfer.

Unifying Theories of Programming~(UTP) is Hoare \& He's long-term research agenda~\citep{HoareH1998}. Their intention is to explore a common basis for understanding the semantics of the modelling notations and programming languages used in describing the behaviour of computer-based systems. Their technique is to describe diverse modelling and programming paradigms in a common semantic setting. They isolate the individual features of these paradigms to emphasise commonalities and differences. They devise formal, often approximate, links between theories to translate predicates from one theory into another. The links also translate specifications into designs and programs as a development method. Understanding the links between formal methods is important, especially for building tool chains for heterogeneous approaches.
 
Beyond the bottom-up construction of tool chains, the
\textsc{AutoFOCUS}
project\footnote{\url{https://www.fortiss.org/ergebnisse/software/autofocus-3}}
\citep{Huber1996-AutoFocustooldistributed,Broy2010-LogicalBasisComponent}
is an example of a long-term effort to provide a formally based
seamless specification, modelling, and development environment, with
methodological support from requirements capture down to code
generation, testing, and artefact evolution.  Several large case studies in model-based development of embedded software were conducted over the years using different \textsc{AutoFOCUS} generations.

The profit from FMs is supposed to be maximal, when thoroughly integrated in a company's design and verification processes \citep{Gleirscher2019-NewOpportunitiesIntegrated}. The chip industry was one of the first sectors where (automated) theorem provers and model checkers have been routinely applied to scrutinize their ever more complex circuits, for instance at INTEL~\citep{DBLP:conf/lics/Harrison03,DBLP:conf/spin/Fix08}, IBM~\citep{DBLP:journals/fmsd/Ben-DavidEGW03} and Oracle~\citep{7886673}.
Perhaps this is due to the fact that chips are mass produced, hence the costs of errors are high, thus the effort of applying FMs paid off early.

Another traditional sector for the application of formal methods is
the railway signalling domain, which can be easily explained by their
safety-critical nature. Very early applications of formal methods to railways
have been reported~\citep{DBLP:journals/corr/abs-2107-05413}. Many
European projects (e.g., FMERail, INESS) and indeed whole conferences
(e.g.,
RSSRail\footnote{\url{https://rssrail2022.univ-gustave-eiffel.fr}})
studied the application of such methods to the railway domain. 
Although this could still be an academic exercise, increasingly the
agenda of formal methods in railways is set by engineering companies (SHIFT2RAIL\footnote{\url{https://shift2rail.org}}) and infrastructure managers (EULYNX\footnote{\url{https://www.eulynx.eu}}). Indeed, the latter are fastly building up expertise centers in model based software engineering and formal verification. 

In the past, a successful route to the wider deployment of formal methods in practice has been the standardisation of their notations, for example, through ISO.  Notable standardisation efforts in this regard are, for instance, LOTOS \cite{ISO8807}, SDL \cite{ITU-SDL-2010}, and the Z notation \cite{ISO-IEC-13568-2002}.

Finally, formal methods are now also applied routinely in purely software-based platforms. An important initial example was the SLAM project at Microsoft \citep{DBLP:journals/cacm/BallLR11}, aimed at Windows device driver compliance. Also Facebook~\citep{OHearn2018,DBLP:journals/cacm/DistefanoFLO19} %
and Amazon Web Services~\citep{DBLP:journals/cacm/NewcombeRZMBD15,DBLP:journals/spe/ChongCEKKMSTTT21} have reported on the application of formal methods for their infrastructure at a massive scale. 
Perhaps, this happened because FMs have matured. Another possible explanation is that the availability and security requirements to contemporary software platforms are extremely high.
These platforms have taken up the role of critical infrastructure.
Other highlights in verified software are the formally verified optimizing compiler CompCert~\citep{DBLP:journals/cacm/Leroy09} and the formally verified Operating System Microkernel seL4~\citep{DBLP:journals/cacm/HeiserKA20}.

It could be argued that an even wider adoption can only be realized by making FM available to average software engineers, who have received a MSc degree in computing or engineering. Apart from professional, easy to use tool support, this requires
insight in the trade-off between investments in and benefits from FM application, as advocated in~\cite{Gorm1}. It also requires an integration of FM
tools with other artefacts in the usual design processes, for instance in agile development~\cite{Gorm2}.

The evidence available from these success stories ranges from single to aggregated 
 opinions of experts as well as anecdotal to very systematic case studies and thorough yet sporadic tool evaluations.
However, data from across a representative range of samples has hardly ever been rigorously measured (e.g., using controlled method experiments).
Hence, albeit impressive, this evidence is still insufficient to underpin a strong argument for a wider deployment of formal methods in industry.  And without such a deployment, further FM research is at risk of getting inapplicable.

\section{An Impact-oriented Plan for Actions}
\label{sec:implications}

A manifesto should of course follow a certain aim, suggest possible actions, disclose
the various impacts hoped for, and discuss relevant implications.

\paragraph{Overview of Expected Impacts of the Manifesto.}
\label{sec:expect-impacts-manif}

We expect a manifesto on applicable formal methods to:
\begin{enumerate}
\item Foster the \textbf{collection (and curation) of} real (small, medium, large)
  \textbf{open problems} (inspired by the success stories in
  Section~\ref{sec:success-stories}) to be tackled by formal methods.  At the lowest level,
  these can be benchmarks defined by practitioners, formal method users, or
  regulators (e.g.,~a ``FM with industry week'' with short-term
  interactions to identify problems at a national or international
  level and follow-up commitments).
\item Provide \textbf{guidance on} how to perform formal method
  \textbf{case studies} and write case study papers and how to review
  them.  We define a case study as an intensive examination of a single example with an aim to generalise across a larger set of examples. It's this generalisation that makes case studies useful in teaching and in industrial practice.
\item \textbf{Stimulate new research proposals} and interdisciplinary research
  collaboration, for example, to improve the
  interface between different formal methods and their users (e.g.,~increase trust through
  \emph{Explainability}, see Table~\ref{tab:principles}), to investigate the economic benefits through
  formal methods (e.g.,~economical value, metrics), or to develop new business models
  integrating such methods.
\item \textbf{Strengthen the community} of researchers that
  \begin{inparaenum}[(i)]
  \item perform evaluations of existing formal approaches and new variants
    in practical contexts,
  \item develop new formal approaches with an interest on achieving
    applicability early, and
  \item support the transfer of these methods into dependable systems
    practice.
  \end{inparaenum}
\end{enumerate}
We detail some of these impact categories below.

\paragraph{Impact on the Conduct, Writing, and Review of Formal Method Research.}

Showing the novelty of research on applicable formal methods w.r.t.~the state of the art is more complicated than showing the novelty of a particular formal technique. A formalism and its expressive power can be explained by examples and the superiority of an algorithm or tool
can be demonstrated by experiments, e.g.,~comparing a range of settings.
However, the \emph{evaluation of the applicability of a formal method as a whole} is more intricate.  So, what is the recommended way for research on applicable formal methods? How can one demonstrate its novelty w.r.t.~the state of the art?  

A few (old) answers \citep{Basili1986-Experimentationsoftwareengineering} within software engineering research are: case studies \citep{Yin2013-CaseStudyResearch}, action research \citep[Sec.~5.5]{Shull2008-GuideAdvancedEmpirical}
and controlled method experiments
\citep{Wohlin2012-ExperimentationSoftwareEngineering}
\citep[Ch.~8]{Shull2008-GuideAdvancedEmpirical}. Following these
methods would greatly benefit the FM community;
Yin's\footnote{I.e.~plan, design, prepare, collect, analyse, and share.} and Wohlin et al.'s procedures provide welcome guidance.  Specific guidelines will effectively aid researchers in conducting evaluation research, writing up results, and performing peer reviews in a repeatable, standardised, and fair manner.
The principles of the manifesto~(Section~\ref{sec:princ-appl-form})
may serve as an initial template for such guidelines.

\paragraph{Implications of the Manifesto on Future Formal Method Teaching.}
\label{sec:impl-manif-future}
It is important to have good case studies that are relevant to students. They must be able to recognise the problem being solved. They should have realistic case studies for every important concept in the course. This is particularly important for industrial courses, where it helps if the presenter has good industrial experience using the formal method. Robust tools are important. There must be parsers and type checkers. Model checkers are attractive, but can disappoint if newcomers have difficulty in scaling their use. Theorem provers have a higher entry barrier but their success can be inspiring. A successful course teaches not just one formal method, but families of formal methods: students like to see the connections between different formal methods. Industrial courses should show how formal methods fit into software management processes and popular methodologies. This includes combining formal methods with testing strategies and their role in formal domain engineering as part of requirements engineering.

\paragraph{Impact on the Evaluation of Future Formal Method Research.}

We expect the manifesto to motivate researchers to carry out
\begin{inparaitem}[]
\item comparative method and tool evaluations~(e.g., \cite{Ferrari2021-SystematicEvaluationUsability}),
\item realistic case studies and goal-directed action research, and
\item controlled method experiments improving over previous lessons learnt \citep{Pfleeger1997-Investigatinginfluenceformal,Sobel2002}.
\end{inparaitem}
For example, in the ABZ community there are ongoing activities to create a case study library for such purposes.\footnote{\url{https://abz2021.uni-ulm.de/case-study}}
Another example is the \emph{VerifyThis} collaborative long-term verification challenge bringing together FM researchers to show ``that deductive program verification can produce relevant results for real systems with acceptable effort''.\footnote{\url{https://verifythis.github.io}}

The manifesto has the potential to create new lines and formats of research funding specifically shaped to the needs of formal method evaluation and tool development, such as funding for experiments and  entrepreneurship funding for spin-offs.  Comparative method experiments and usable tool interfaces require resources going beyond PhD projects or beyond the pure response to scientific questions.  Only appropriately funded research projects will create convincing evidence.

\paragraph{Impact on the Further Development of the Formal Methods
  Community.}

The manifesto could reduce the current fragmentation of the formal
methods community by subsequently integrating selective
sub-communities, for example, communities working on common semantic
frameworks (e.g.,~the UTP community\footnote{\url{https://www.cs.york.ac.uk/circus/utp2019}}) or formal method integration (e.g.,~the
sub-communities around the ``Formal Methods in Industrial Critical Systems (FMICS)", 
``Integrated Formal Methods (iFM)'', ``NASA Formal Methods (NFM)'', and ``Software Engineering and Formal Methods (SEFM)'' conference
series\footnote{\url{http://fmics.inria.fr}, \url{http://www.ifmconference.org}, \url{https://shemesh.larc.nasa.gov/nfm2021}, \url{https://sefm-conference.github.io}}).  Moreover, the
manifesto could inspire new actions of researchers to work towards a
collection of formal methods that follow the proposed principles.

\paragraph{Impact on Software Engineering as a Legally Recognised
  Profession.}

In his Turing Award acceptance speech about 40 years ago, Tony Hoare
reviewed type safety precautions in programming languages and
concluded: ``In any respectable branch of engineering, failure to
observe such elementary precautions would have long been against the
law'' \citep{Hoare1981-EmperorsOldClothes}.  In this regard, for
example, U.S.~law still does not recognise computing (including
software engineering) as a profession
\citep{Choi2021-Softwareprofessionalsmalpractice}, opposing ACM's
self-perception \citep{Chien2017-Computingisprofession}.  This is mainly because software practitioners' work is
not subject to malpractice claims based on a legal concept known
as ``customary care''.  Customary care defines
(i.e.,~standardises) best practice more stringently\footnote{Taking the state of the art including recent scientific results as a reference.} than the notion 
of ``reasonable care'' applied to any occupation or business.  From a computing standpoint,
ongoing juristic debates about which other
occupations\footnote{``[E]very court to consider this question has
  refused to recognize software developers as professionals''
  \citep[p.~23]{Choi2021-Softwareprofessionalsmalpractice}.} should be
treated as professions could be advanced by this manifesto,
corroborate codes such as ACM's Code of Ethics and Professional
Conduct,\footnote{Version~2018: \url{https://www.acm.org/code-of-ethics}}
and help to standardise results from the formal methods community as
credible best practices underpinning such codes.

In Denmark, 51\% of the developers hired by the IT companies developing software do not have a BSc/MSc degree in computing.\footnote{\url{https://www.prosa.dk/artikel/nu-er-der-over-100000-it-professionelle}}
It is to be expected that this situation can be generalised to other European countries and, to a smaller degree, also to critical application domains.  Hence, this manifesto could aid in the expansion of existing software 
engineering professionalism\footnote{\url{https://en.wikipedia.org/wiki/Software_engineering_professionalism}} within such domains.

\section{A Life Without the Manifesto}
\label{sec:inaction}

Above, we summarised actions and expected outcomes of a successful
implementation of the manifesto.  However, some negative long-term
consequences of not following an agenda implied by the manifesto are
to be foreseen.

First, the progress of formal method research might be further
threatened by missing scalability, vacuous proofs, lack of user
education and training, poor tool integration, lack of researcher
engagement and, thus, research funding
\citep[pp.~117:23,29]{Gleirscher2019-NewOpportunitiesIntegrated}.

Secondly, formal methods might be wiped out by opportunistic trends or
powerful convenience technologies (e.g.,~relying too much on search-
or AI-based software engineering) that can make the highlighted
problems worse.  It can be observed that software solutions
constructed through automatic search may require significant further
investments into the reverse engineering of these solutions in order
to verify them.  This may happen frequently in cases where not all
critical properties to be verified can be encoded into the search
criteria.

Ultimately, decreasing global coordination among formal method researchers can lead
to an extinction of the formal methods community, which is currently
rather fragmented. %
It is difficult for the community to maintain too
many notations and too many tools and make fast progress.  This
situation seems quite unique
among related or other scientific disciplines (i.e.,~STEM\footnote{Science,
  Technology, Engineering, Mathematics}).  Also, there is a proliferation of
formal method conferences and workshops that are competing for the same resources
(i.e.,~papers, reviewers, etc.). Ideally, a representative, coordinated
approach could lead to an authoritative voice towards the scientific,
governmental, and industrial communities.

\section{Conclusions and Outlook}
\label{sec:conclusions}

The manifesto for applicable formal methods expresses aims and
intentions and shall help formal methods researchers to implement a
modern research agenda for developing formal methods that can arguably
be used for critical software engineering research but, even more
importantly, in the practical engineering of systems and software
whose functioning is critical and whose failure would have
unacceptable consequences.  Rather than exercising criticism of past
developments, the manifesto strives to foster progress of a currently
dissatisfying situation found in the science of formal methods.

\small
\bibliography{}
\end{document}